\documentclass[sigconf,screen,nonacm]{acmart}

\usepackage{tabularx}
\usepackage{hyperref}
\usepackage{microtype}
\usepackage{balance}

\usepackage{xeCJK} 

\begin{document}
\begin{sloppy}
\title[Operationalizing Ethics for AI Agents: How Developers Encode Values into Repository Context Files]{Operationalizing Ethics for AI Agents:\\
How Developers Encode Values into Repository Context Files}

\author{Christoph Treude}
\orcid{0000-0002-6919-2149}
\affiliation{%
  \institution{Singapore Management University}
  \city{Singapore}
  \country{Singapore}
}
\email{ctreude@smu.edu.sg}

\author{Sebastian Baltes}
\orcid{0000-0002-2442-7522}
\affiliation{%
  \institution{Ruprecht-Karls-Universität Heidelberg}
  \city{Heidelberg}
  \country{Germany}
}
\email{sebastian.baltes@uni-heidelberg.de}

\author{Marc Cheong}
\orcid{0000-0002-0637-3436}
\affiliation{%
  \institution{University of Melbourne}
  \city{Melbourne}
  \country{Australia}
}
\email{marc.cheong@unimelb.edu.au}

\renewcommand{\shortauthors}{Treude et al.}

\begin{abstract}
As AI coding agents become embedded in software development workflows, developers are beginning to operationalize ethical principles by encoding behavioral rules into repository-level context files for AI agents, such as \texttt{AGENTS.md} files. Rather than examining the ethics \emph{of} AI agents in the abstract, this vision paper investigates how ethics and values are already being translated \emph{for} AI agents into actionable instructions that shape agent behavior. Through a preliminary investigation, we find that developers are already embedding guidance related to fairness, accessibility, sustainability, tone, and privacy. These artifacts function as a developer-authored governance layer, translating abstract principles into situated, natural-language directives within development workflows. We outline a research agenda for studying this emerging practice, including how encoded values vary across communities, what governance dynamics emerge when multiple contributors negotiate these files, and whether agents reliably adhere to the constraints specified.  Understanding how ethics and values are operationalized \emph{for} AI agents is essential to ground AI governance in modern software engineering practice.
\end{abstract}



\maketitle

\section{Introduction}

Ethical principles for AI systems, such as fairness, accountability, transparency, and safety, are widely discussed in research and policy. However, translating these principles into everyday engineering practice remains difficult. Principles might be too vague, unenforceable, or downright clash with each other \cite{Cheong2025-ng}. Engineers may struggle to translate ethical principles into design requirements.

The increasing integration of AI coding agents into development workflows introduces a new dimension to this long-standing challenge. These agents operate with varying degrees of autonomy and human involvement. They generate code, propose refactorings, review pull requests, and interact directly with repository artifacts. Their behavior is shaped not only by model training and prompts, but also by contextual artifacts embedded in repositories.

In this paper, we argue that a new and largely unexplored phenomenon is emerging: developers are operationalizing ethics \emph{for} AI agents by encoding behavioral constraints directly into repository-level context files. Files such as \texttt{AGENTS.md} contain instructions that specify how AI agents should behave within a project. These instructions are written by humans, interpreted by machines, and embedded in development workflows.

Prior work has explored ethics-by-design, embedded governance mechanisms, and Responsible AI engineering processes as ways to integrate ethical principles into development practice~\cite{friedman2019value, spiekermann2023value}. These efforts have primarily targeted human developers through organizational policies and documentation artifacts. Context files introduce a distinct operational layer: machine-interpretable constraints embedded directly in repositories and interpreted by AI agents during development workflows. Ethical frameworks have never been panaceas in human endeavors; practitioners regularly face conundrums that principles alone cannot resolve. AI agents working under natural-language repository directives will face analogous complexity. This translation layer between abstract principles and concrete agent behavior constitutes a new object of study for software engineering research.

\section{Background and Related Work}

\begin{table*}[!t]
\caption{Examples of Ethical Operationalization in Context Files: Excerpts from Six Open-Source Repositories.}
\label{tab:examples}
\centering
\begin{tabularx}{\textwidth}{p{4cm} X}
\toprule
\textbf{Link} & \textbf{Verbatim Excerpt} \\
\midrule

\href{https://github.com/Abhiek187/ez-recipes-web/blob/main/AGENTS.md}{Abhiek187/ez-recipes-web} &
``When working with the user, ensure you follow all guidelines for ethical AI, such as keeping the human in the loop, taking accountability for changes, and being transparent\ldots'' \\

\addlinespace

\href{https://github.com/github/awesome-copilot/blob/main/agents/se-responsible-ai-code.agent.md}{github/awesome-copilot} &
``Build systems that are accessible, ethical, and fair. Test for bias\ldots'' [...]
\texttt{test\_names = ['John Smith', 'José García', 'Lakshmi Patel', 'Ahmed Hassan', '李明']} [...]
``Different outcomes for same qualifications but different names'' \\

\addlinespace

\href{https://github.com/D7460N/DHCP/blob/main/AGENTS.md}{D7460N/DHCP} &
``Explicitly Avoid \ldots Moral lectures or unsolicited opinions'' \\

\addlinespace

\href{https://github.com/haxtheweb/create/blob/main/AGENTS.md}{haxtheweb/create} &
``Accessible: HAX maximizes accessibility\ldots'' [...]
``Sustainable: Environmental (less data, lower battery usage)\ldots and community\ldots sustainability.'' \\

\addlinespace

\href{https://github.com/home-assistant/core/blob/dev/AGENTS.md}{home-assistant/core} &
``Inclusivity: Use objective, non-discriminatory language'' [...]
``Clarity: Write for non-native English speakers'' \\

\addlinespace

\href{https://github.com/tmobile/magentaA11y/blob/main/AGENTS.md}{tmobile/magentaA11y} &
``Respectful, Inclusive Language\ldots'' [...]
``Bias-Aware and Error-Resistant\ldots'' [...]
``Verification-Oriented Responses\ldots'' \\

\bottomrule
\end{tabularx}
\end{table*}

Our work draws on two bodies of research: the integration of ethics and values into software engineering practice, and the study of AI agents in software development. The first informs what ethical principles might be encoded in repository artifacts and how; the second informs where such encoding occurs and what agents do with it.

\subsection{Ethics, values, and SE}
Ethics and values have long been discussed in software engineering and in digital ethics, \textit{writ large}. Foundational efforts such as the \emph{ACM/IEEE Software Engineering Code of Ethics} articulate professional obligations and societal responsibilities of software engineers~\cite{gotterbarn1997software}, while various other frameworks and principles articulate the same on a broader level~\cite{Hagendorff2020-tk, Cheong2025-ng}. However, several authors argue that such codes provide limited actionable guidance for everyday design decisions and must be complemented by explicit ethical deliberation within development teams~\cite{gogoll2021ethics, lurie2016professional}, resonant with critiques of applied AI/digital/technological ethics ``...in general... lack[ing] mechanisms to reinforce its own normative claims'' \cite{Hagendorff2020-tk}.

Beyond professional conduct and on a broad practice- and policy-level, research has explored how ethical values can be integrated into software processes. Value-Sensitive Design~\cite{friedman2019value} and Value-Based Engineering~\cite{spiekermann2023value} aim to make human values explicit in system design, while the IEEE 7000 standard proposes structured process models to address ethical concerns during system development~\cite{spiekermann2021expect}. Systematic reviews show growing interest in operationalizing human values and acceptable norms \cite{Gao2024-gi} in software engineering, particularly in the requirements and design phases, but reveal limited support for later lifecycle stages and implementation practices~\cite{shahin2022operationalizing, alidoosti2022ethics}. 

Several approaches translate ethical principles into concrete development artifacts. The Responsible AI Pattern Catalogue~\cite{lu2024responsible} presents governance, process, and product patterns that translate ethical principles into system-level practices, including standardized reporting templates, ethical requirements, and governance structures intended for human developers and organizations. \emph{Ethical User Stories} and related agile practices embed concerns such as fairness, accessibility, and sustainability into backlog items and sprint routines~\cite{halme2022ethical, zimmermann2024bringing}. Goal-oriented methods derive \emph{Social, Legal, Ethical, Empathetic, and Cultural} (SLEEC) requirements from explicit value models~\cite{junior2026operationalizinghumanvaluesrequirements}, while recent work advocates for lightweight, proactive integration of such considerations into existing engineering workflows~\cite{kapferer2024towards}.

\subsection{Agents and Agents4SE}
Studies on \textit{in silico} agents, namely the increasing need for ethical conduct in their use, design, deployment, and emergent behaviors, are not new concepts. Consider traditional agent-based social simulations: ethical issues are found to ``...arise from both its practice and its organisation''~\cite{anzola2022}. This `top-down approach' (from the practitioner's perspective) is also complemented by a similar call from the `bottom-up approach' (from the agent's perspective), such as the operationalization of `good' versus `bad' behavior at agentic level~\cite{Mascaro2010-cg}. Most pre-date modern generative AI.

More recent work focuses on AI coding agents specifically. A taxonomy of human--AI collaboration in software engineering characterizes how developers interact with AI tools across different roles and intensities~\cite{treude2025developers}. CRAFT (comprehensive, responsible, adaptive, foundational, translational) has been proposed as a set of values for agentic SE that extends the focus of agent design beyond coding to human--AI collaboration~\cite{hoda2025toward}. Building on this framing, recent work identifies trustworthiness dimensions for AI software engineers spanning technical quality, transparency and accountability, epistemic humility, and societal and ethical alignment~\cite{aleti2026trustworthyaisoftwareengineers}. This body of work, however, addresses agent capabilities, evaluation, and proper conduct (\textit{vis-`a-vis} their deployers) from the `bottom-up' rather than how developers encode persistent behavioral constraints within repositories.

Recent research has begun to conceptualize repository-level context files as a distinct class of artifacts in agentic software development. An analysis of 2{,}303 such files (``Agent READMEs'') across 1{,}925 repositories finds that they resemble configuration code, primarily encoding build instructions, implementation details, and architectural information, while security and performance considerations are rarely specified~\cite{chatlatanagulchai2025agent}. Another study examining 466 open-source projects reports substantial variation in how context is expressed (descriptive, prescriptive, prohibitive, explanatory, conditional), without an established content structure~\cite{mohsenimofidi2026context}. A broader study covering eight configuration mechanisms across 2{,}853 repositories identifies \texttt{AGENTS.md} as an emerging interoperable standard across tools~\cite{galster2026configuring}. Empirical evaluations further suggest that \texttt{AGENTS.md} can lower agent runtime and token consumption without degrading task completion~\cite{lulla2026impact}, and the format is used in over 60{,}000 open-source projects.\footnote{\url{https://agents.md/}
, accessed 2026-05-06.}

Across these bodies of work, prior research has examined professional ethics, value-based development processes, agent trustworthiness, accountability and governance of AI coding tools~\cite{treude2026accountable}, considerations for ethics in agent-based studies, and the structure and efficiency implications of context files. While existing Responsible AI frameworks translate ethical principles into governance processes and documentation artifacts for human developers~\cite{lu2024responsible}, context files target AI agents directly. How ethical principles are translated into such machine-interpretable artifacts remains largely unexplored, and we address this gap by treating \texttt{AGENTS.md} as a concrete operational layer for encoding ethics in agentic software development.

\section{Preliminary Investigation}

To explore how developers operationalize ethics \emph{for} AI agents, we used a combination of GitHub code search and ChatGPT web search to identify repositories containing \texttt{AGENTS.md} files with potential ethics-related guidance. We then manually inspected 25 repositories and selected six illustrative examples, presented in Table~\ref{tab:examples}, to motivate the research agenda rather than to offer a comprehensive empirical account. Even in this small exploratory sample, ethical principles are not merely referenced but translated into concrete, machine-interpretable constraints.

For example, one repository instructs the agent to ``follow all guidelines for ethical AI'', explicitly emphasizing keeping ``the human in the loop'', ``taking accountability for changes'', and being ``transparent''. While this appears to translate abstract notions of oversight into interaction-level constraints, it remains unclear what concrete behavioral changes an AI agent could derive from such high-level and ambiguous directives. The instruction documents ethical intent, but it does not specify executable conditions, triggers, or enforcement mechanisms. In contrast, another repository operationalizes fairness through structured bias testing: beyond requiring systems to be ``accessible, ethical, and fair'', it provides explicit test data such as \texttt{['John Smith', 'José García', 'Lakshmi Patel', \ldots]} and directs the agent to check for ``different outcomes for same qualifications but different names'', embedding fairness as executable evaluation logic rather than a high-level aspiration. Other repositories encode normative boundaries on agent behavior, for instance by instructing agents to ``avoid moral lectures or unsolicited opinions'', or by mandating ``objective, non-discriminatory language'' and communication suitable for non-native English speakers. Across these examples, accountability, fairness, inclusivity, and tone become machine-interpretable constraints intended to shape agent behavior.

These examples demonstrate that developers are actively selecting the ethical concerns that matter in their projects and reformulating them as machine-interpretable constraints. Fairness becomes bias testing logic, accountability becomes interaction constraints, inclusivity becomes linguistic guidance, and sustainability becomes a design requirement. Ethical commitments are therefore not simply declared at the level of principles; they are embedded directly into the instructions that are meant to shape agent behavior within development workflows. This practice is still nascent, and key questions remain: given that humans grapple with the distinction between `doing the right thing' and `doing things right', it is unclear how AI agents can be expected to handle such distinctions when guided only by natural-language repository directives.

\section{Roadmap}

The emergence of repository-level context files that encode behavioral constraints for AI agents has introduced a new dimension to discussions of ethics and AI: rather than debating the ethics \emph{of} AI agents in the abstract, developers are beginning to operationalize ethics \emph{for} AI agents within everyday development workflows. This development opens up a new research frontier for software engineering. A first step is large-scale empirical mapping of context files to identify which ethical categories are encoded most frequently and which remain absent. This work can reveal whether bias, accessibility, sustainability, or privacy dominate developer attention and how these emphases vary across domains and cultures. This will also reveal which value choices developers prioritize when instructing their agents, in turn reflecting the diversity of values humans uphold.

Beyond categorization, it is important to study the translation process itself. How do developers decide which ethical principles to encode? How broad (or specific) are these ethical directives in such files? How are these constraints negotiated in pull requests? How are these principles negotiated between human software engineers and AI counterparts? Mining repository histories combined with qualitative analysis can uncover the socio-technical dynamics underlying ethical operationalization.

Equally important is evaluating whether agents adhere to encoded constraints. Future work should experimentally compare agent behavior with and without context files, measuring compliance with constraints related to bias mitigation, tone, accessibility, and regulatory guidance. Such studies would show whether operationalization shapes agent behavior or merely signals intent.

Longitudinal analysis can further examine how `encoded ethics' evolves. Ethical commitments and directives can be strengthened after incidents, relaxed under productivity pressure, or adapted to new regulations. Some might be an exercise in `box-ticking', while others are genuine commitments to ethics, with varying degrees of practice, paralleling the trend of ethical documentation for humans \cite{Gao2024-gi}. Repository histories make it possible to study how such shifts unfold over time.

At the same time, it remains unclear whether how ethics are communicated to AI agents should resemble the way ethics are traditionally documented for human contributors, such as in codes of conduct or policy statements \cite{tourani2017code}. Human-oriented ethical documentation often relies on shared norms, contextual judgment, and implicit understanding. Determining whether similar forms of documentation suffice or whether fundamentally different, machine-oriented representations are required is an important direction for future research.

Future work should also look beyond natural-language context files to the full range of agent configuration artifacts. Agent skills, tool permission settings, hook scripts, and system prompt files each constitute distinct configuration layers that may encode ethical constraints in forms quite different from prose directives~\cite{galster2026configuring}. Whether these programmatic artifacts reproduce, complement, or contradict constraints found in natural-language files is unknown; a complete account of operationalized ethics must consider the entire configuration space agents operate within.

Developers are already operationalizing ethics \emph{for} AI agents in current repositories. By studying how these operational constraints are encoded, interpreted, and revised, software engineering research can move beyond abstract principles to an empirical understanding of how AI governance is implemented in practice. Moving forward, these insights can be applied beyond SE to include AI agent behavior in other domains, such as business and social sciences.

\begin{acks}
We thank the organizers and participants of the scoping workshop ``AI in a Fragmented World: Navigating Trade-offs Across Disciplines and Cultures'', held at the Speinshart Scientific Center for AI and SuperTech (SSC) in February 2026, for discussions that shaped this work. Marc would like to acknowledge the institutional support provided by Google (as part of the \textit{Philosophical, tech, and legal perspectives on human relationships with social robots} project at CAIDE) for funding his travel to SSC.
\end{acks}

\balance

\end{sloppy}
\end{document}